\documentclass[10pt,conference]{IEEEtran}
\IEEEoverridecommandlockouts
\usepackage{cite}
\usepackage{amsmath,amssymb,amsfonts}
\usepackage{subfig}
\usepackage{subfloat}
\usepackage{tabularx}
\usepackage{array}
\usepackage{algorithm}
\usepackage[noend]{algpseudocode}
\usepackage{stfloats}
\usepackage{graphicx}
\usepackage{textcomp}
\usepackage{booktabs}
\usepackage{balance}
\usepackage{subcaption}
\def\BibTeX{{\rm B\kern-.05em{\sc i\kern-.025em b}\kern-.08em
    T\kern-.1667em\lower.7ex\hbox{E}\kern-.125emX}}
    
\newcommand{\solution}{\textit{Voyager}}

\newcommand{\ie}{\textit{i.e., }}  

\def \1{\textit{(i)}}
\def \2{\textit{(ii)}}
\def \3{\textit{(iii)}}
\def \4{\textit{(iv)}}
\def \5{\textit{(v)}}

\begin{document}
\author{
    \IEEEauthorblockN{Chao Feng\IEEEauthorrefmark{1}, Alberto Huertas Celdrán\IEEEauthorrefmark{1}, Michael Vuong\IEEEauthorrefmark{1}, G\'er\^ome Bovet\IEEEauthorrefmark{2},
    Burkhard Stiller\IEEEauthorrefmark{1}}
    
    \IEEEauthorblockA{\IEEEauthorrefmark{1}Communication Systems Group CSG, Department of Informatics, University of Zurich UZH, CH--8050 Zürich, Switzerland \\{[cfeng, huertas, stiller]}@ifi.uzh.ch, michael.vuong@uzh.ch}
    \IEEEauthorblockA{\IEEEauthorrefmark{2}Cyber-Defence Campus, armasuisse Science \& Technology, CH--3602 Thun, Switzerland gerome.bovet@armasuisse.ch}
}
\DeclareRobustCommand*{\IEEEauthorrefmark}[1]{%
  \raisebox{0pt}[0pt][0pt]{\textsuperscript{\footnotesize #1}}%
}

\title{\solution{}: MTD-Based Aggregation Protocol for Mitigating Poisoning Attacks on DFL}
\maketitle

\begin{abstract}

The growing concern over malicious attacks targeting the robustness of both Centralized and Decentralized Federated Learning (FL) necessitates novel defensive strategies. In contrast to the centralized approach, Decentralized FL (DFL) has the advantage of utilizing network topology and local dataset information, enabling the exploration of Moving Target Defense (MTD) based approaches. 

This work presents a theoretical analysis of the influence of network topology on the robustness of DFL models. Drawing inspiration from these findings, a three-stage MTD-based aggregation protocol, called \solution{}, is proposed to improve the robustness of DFL models against poisoning attacks by manipulating network topology connectivity. \solution{} has three main components: an anomaly detector, a network topology explorer, and a connection deployer. When an abnormal model is detected in the network, the topology explorer responds strategically by forming connections with more trustworthy participants to secure the model. Experimental evaluations show that \solution{} effectively mitigates various poisoning attacks without imposing significant resource and computational burdens on participants. These findings highlight the proposed reactive MTD as a potent defense mechanism in the context of DFL.

\end{abstract}


\section{Introduction}
\label{intro}
The widespread adoption of communication technologies and the growing prevalence of the Internet-of-Things (IoT) concept are leading to an extensive rise in the number of devices connected to the Internet. However, these massive numbers of devices create enormous amounts of distributed data, posing new challenges in the processing and analysis of such data in a privacy-preserved way. Federated Learning (FL) is a novel approach in that leverages collaborative learning on distributed datasets~\cite{mcmahan2016federated}. 
Unlike traditional Machine Learning (ML) methods, FL allows users to share their local models instead of the raw data with a central server, preserving data privacy. However,  a central server is still needed for FL to aggregate the models. This Centralized FL (CFL) approach has some drawbacks, such as the single point of failure and network bottlenecks. To address these concerns, Decentralized FL (DFL) has been introduced, where all nodes are considered equal and can act as both trainers and aggregators simultaneously, improving the fault tolerance of the system~\cite{beltran2023decentralized}.

Nevertheless, due to its distributed nature, FL is vulnerable to adversarial attacks whereby malicious participants can undermine the privacy and robustness of the FL model~\cite{bouacida2021vulnerabilities}. Poisoning attacks constitute a category of malicious attacks that exploit the robustness and precision of FL. Malicious participants manipulate their own data or models and share them with other participants to undermine the robustness of their models. Poisoning attacks pose significant risks to both CFL and DFL. In CFL, the central server receives poisoned models from malicious participants, leading to a compromised global model after aggregation. In DFL, the poisoned models produced by malicious participants are disseminated throughout the network, presenting a threat to all participants involved.


In contrast to CFL, DFL possesses the capacity to leverage supplementary information, including network topology and local datasets, thereby enabling the exploration of novel approaches in devising defense strategies. Introduced in 2009, Moving Target Defense (MTD) is a revolutionary security approach that aims to mitigate the impacts of cyberattacks by actively or passively altering specific system parameters, including network configurations, file systems, or system libraries \cite{cai2016moving}. Through the analysis of the functioning of DFL systems, this work demonstrates the impact of network topology on the security vulnerability of DFL from a theoretical standpoint. Specifically, for participants involved in a DFL network, the number of connections to malicious participants follows a hypergeometric distribution, which indicates the risk of being connected with a malicious participant has a positive correlation with the average number of connection with other neighbors as well as the proportion of malicious participants, and a negative correlation with the number of benign neighbors. Consequently, this work establishes the feasibility of implementing MTD-based defensive measures through modifications to the network topology.


Inspired by the theoretical discoveries, this work proposes an MTD-based aggregation protocol, named \solution{}, which reactively manipulates the topology of the network to enhance the robustness of the DFL  system. Specifically, the \solution{} protocol consists of three stages: an anomaly detector, a network topology explorer, and a connection deployer. The anomaly detector compares the local model with models shared by other participants to determine if an anomalous model has been received, thus triggering an MTD policy. Once an anomaly is detected, a triggering message is sent and the network topology explorer attempts to identify more trustworthy participants by adding them to the candidate set. Finally, the connection deployer is responsible for establishing connections with these candidate participants and sharing the respective models to create an aggregated model. 

Therefore, the contributions of this work encompass the following: \1 Modeling and analyzing the impact of network topology on the security risk of DFL, and identifying the primary factors affecting the robustness of DFL; \2 Proposing a reactive MTD-based aggregation protocol, referred to as \solution{}, based on the theoretical analysis, to mitigate the impact of poisoning attacks on DFL; \3 Developing a prototype of \solution{}, with publicly available in \cite{Voyager}, within the FedStellar~\cite{beltran2023fedstellar} DFL framework; \4 Conducting a series of experiments to validate the performance of \solution{}. These experiments involve assessing various network topologies (such as ring, star, random, and fully connected networks), employing different attack techniques like label flipping and model poisoning, and evaluating three datasets (MNIST~\cite{lecun_MNISTHandwritten_2010}, FashionMNIST~\cite{xiao_FashionMNISTNovel_2017}, and Cifar10~\cite{krizhevsky_LearningMultiple_2009}) while comparing the results with related works. The experimental results demonstrate that \solution{} effectively reduces different poisoning attacks across various network topologies and datasets, without noticeably increasing the resource consumption and network overhead.


\section{Background \& Related Work}
\label{sec:related}

This section offers an overview of poisoning attacks affecting the robustness of FL systems, as well as a summary of current research on countermeasures against such attacks.
\subsection{Poisoning Attacks}

\begin{figure}[b!]
    \centering
    \includegraphics[width=0.75\linewidth,angle=270, trim={0.5cm 2cm 0.5cm 2cm},clip]{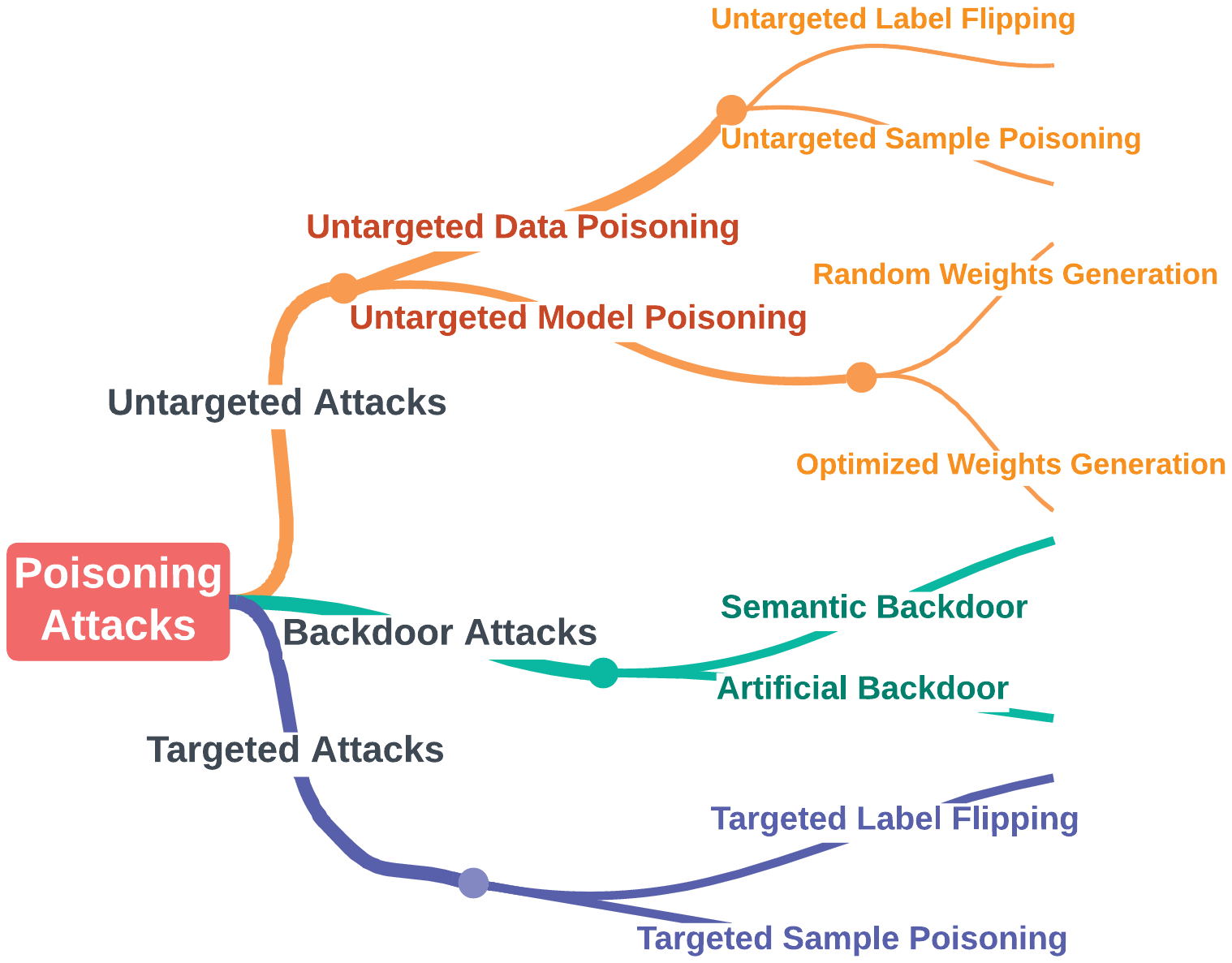}
    \caption{Taxonomy of Poisoning Attacks}
    \label{fig:taxonomy}
\end{figure}

The taxonomy in \figurename~\ref{fig:taxonomy}  provides an overview of poisoning attacks, which can be categorized into three groups based on the intentions of the adversary, namely untargeted, targeted, and backdoor attacks~\cite{tian_ComprehensiveSurvey_2023,xia_PoisoningAttacks_2023}. 

The goal of untargeted attacks is to diminish the overall performance of the model, thereby impeding its convergence~\cite{shejwalkar_BackDrawing_2022}. In instances of targeted attacks, the adversary intentionally alters the training data to induce incorrect or biased predictions from the model on particular target inputs. The objective of the adversary is to infiltrate the model in such a way that only a particular target set or class is misclassified, while the rest of the set is correctly classified~\cite{fung_LimitationsFederated_2020}. The last category is the backdoor attack, where the adversary injects one or multiple triggers into the model during training, which can then be exploited during the inference process~\cite{bagdasaryan_HowBackdoor_2019}. Specifically, the model behaves normally in the absence of the trigger, but an attacker can provoke a desired prediction or classification by presenting the trigger at the inference stage.

From the attack strategy perspective, poisoning attacks can be categorized as data poisoning or model poisoning. Data poisoning involves manipulating training data to influence the model training process indirectly. This can be done by changing labels in the dataset (label flipping) or inserting malicious data samples (sample poisoning). Depending on the attacker's goal, these techniques can be used as untargeted attacks (untargeted label flipping and sample poisoning) or targeted attacks (targeted label flipping and sample poisoning). In the context of backdoor attacks, the attacker can choose to use either semantic or artificial backdoors~\cite{bagdasaryan_HowBackdoor_2019}. Semantic backdoors utilize naturally existing cues as triggers, while artificial backdoors are created by intentionally injecting specific triggers. Model poisoning attacks are more dangerous and harder to detect than other attacks because they allow malicious participants to manipulate the trained model directly~\cite{shejwalkar_BackDrawing_2022}. There are two types of model poisoning attacks: random weights generation, where random noise is added to the transmitted model, and optimized weights generation, where the distribution of the added noise is optimized to decrease the chances of detection.

\subsection{Defense Mechanisms}
In light of the growing concern of poisoning attacks as a serious threat to FL, it is crucial to thoroughly investigate strategies and mechanisms for protecting FL systems from these attacks. The comparison provided in \tablename~\ref{tab:summaryRelatedWork} highlights the most relevant and recent research in this area. Generally, these defense mechanisms could be categorized into three classes based on their approach: robust aggregation rules, secure federated ensemble strategies, and MTD techniques. 

The Krum aggregation rule~\cite{Krum:2017} is designed to identify honest local models from shared models. It uses a deterministic Euclidean distance to compute vectors and is effective when there are more honest nodes than malicious ones. However, Krum has a high risk of selecting the wrong model if more than half of the participants are malicious. Statistical methods play a crucial role in aggregation, with two approaches designed in~\cite{Statistics:2021}, TrimmedMean and Median. TrimmedMean removes extreme values to address outliers, while Median determines the median value from the shared models. Both methods are practical and efficient. The FLTrust~\cite{FLTrust:2021} algorithm incorporates a reputation mechanism where nodes' trustworthiness changes over time based on their performance. Nodes with lower performance have their reputation decrease, while those with higher performance have their reputation increase. Reputation affects the weight of nodes' contributions during aggregation, giving more importance to reputable participants. \cite{Bulyan:2018} introduces Bulyan, a method that improves security against attacks by combining existing byzantine-resilient aggregation rules such as Krum and TrimmedMean. It first uses Krum to identify potentially benign participants and then applies TrimmedMean to aggregate their update parameters individually. This approach aims to protect against malicious workers, with a maximum tolerance of $\frac{b-3}{4}$, where $b$ denotes the number of benign participants. The Scaffold ~\cite{Scaffold} predicts update directions for the server and local models, calculates the difference between them, and corrects the local models based on this difference. This speeds up the convergence process and reduces communication costs but requires the server to store separate parameters for each participant.

\begin{table}[t!]
\caption{Mitigating Attacks for Federated Learning}
\resizebox{\columnwidth}{!}{%
    \begin{tabular}{@{}lllllllll@{}}
    \toprule
    \textit{Reseach}  & \textit{Approach} & \textit{Category} & \textit{\begin{tabular}[c]{@{}l@{}}Network\\ Overhead\end{tabular}} & \textit{Complexity} & \textit{Schema} & \textit{Attack} \\
    \hline
    \cite{Krum:2017}  2017 & Krum & Robust Agg. & Little & Scalable & CFL & Poisoning\\
    \cite{Statistics:2021} 2021 & \begin{tabular}[c]{@{}l@{}}Trimmed\\ Mean\end{tabular}  & Robust Agg. & Little & Scalable & CFL & Poisoning \\
    \cite{Statistics:2021} 2021 & Median  & Robust Agg. & Little & Scalable & CFL & Poisoning \\
    \cite{FLTrust:2021} 2021 & FLTrust & Robust Agg. & Medium & Scalable & CFL  & Poisoning\\
    \cite{Bulyan:2018} 2018 & Bulyan & Robust Agg. & Little & Scalable & CFL & Poisoning\\
    \cite{Scaffold} 2020 & Scaffold & Robust Agg. & Large & Cost heavy  & CFL & Poisoning\\
    \cite{Fed-Ensemble:2021} 2021 & Fed-Ensemble & Ensemble & Small & Medium  & CFL & Poisoning\\
    \cite{Knowledge:2015} 2015 & \begin{tabular}[c]{@{}l@{}}Knowledge\\ Distillation\end{tabular}  & Ensemble & Medium & Scalable & CFL  & Poisoning\\
    \cite{beltrán2023mitigating} 2023 & - & MTD & Small & Scalable & DFL & Comm.\\
    This work & Voyager & MTD & Small & Scalable & DFL & Poisoning\\
        \bottomrule
    \end{tabular}
}
    \label{tab:summaryRelatedWork}
\end{table} 

Ensemble techniques in FL combining multiple models to achieve superior performance by leveraging the collective knowledge of participating models. Fed-Ensemble~\cite{Fed-Ensemble:2021} deviates from the typical FL framework by incorporating multiple independent aggregators. During each iteration, Fed-Ensemble allocates one of these aggregators to chosen participants' subsets, enabling them to conduct training on their respective local datasets. Consequently, this practice enhances performance by leveraging the collective predictions derived from multiple models. Furthermore, this methodology bolsters the robustness of the system against adversarial attacks, as the compromise of a singular model does not exert a substantial influence on the entire system. Knowledge distillation~\cite{Knowledge:2015} utilizes a student-teacher mechanism, where the teacher model, trained on a large dataset for strong generalization, transfers its knowledge to smaller, more practical models through distillation. This process involves creating soft targets, which are probability distributions produced by the teacher model, serving as guides for the student models to learn from the teacher's knowledge. However, ensemble techniques are not Byzantine-robust guaranteed, which is crucial for addressing robustness issues.

However, a limited amount of research delves into the implementation of MTD for safeguarding FL systems, especially for DFL. \cite{beltrán2023mitigating} put forward a proactive MTD-based strategy that involves altering the network configuration and implementing encryption measures to safeguard the DFL system against potential communication vulnerabilities, such as Man in the Middle (MitM) attacks.

To conclude, although numerous methods employ aggregation rules and ensemble strategies, there is currently no research that utilizes the MTD mechanism for the purpose of mitigating poisoning attacks and ensuring the security of FL systems. Furthermore, it is worth noting that all of these aggregation rules and ensemble strategies were originally designed for CFL rather than DFL.

\section{Problem Statement}
\label{sec:problemstatement}
This section presents a formal establishment of the problem, which is essential for developing and evaluating the proposed algorithm. In the $N$ participants federation system, each participant $c_i \in \mathbb{C}$ uses its own local dataset $D_i \in \mathbb{D}$ to train a local model $m_i$, where $\mathbb{C}$ donates the universe of the participants, and $\mathbb{D}$ donates the universe of the dataset. For each data point $(x_{i,j},y_{i,j}) \in D_i$, the loss function associated with this data point is defined as $f(m_i(x_{i,j}); y_{i,j})$. Thus, for the whole local dataset $D_i$ in $c_i$, the corresponding population loss function is formalized as: $F(m_i) = \sum_{(x_{i,j},y_{i,j}) \in D_i}f(m_i(x_{i,j});y_{i,j})$. Hence, the objective of training the local model is to identify the optimal $m_i^\ast$ within the parameter space $\mathbb{M}$, as defined in Equation~\eqref{eq:local_goal}:
\begin{center}
\setlength{\abovedisplayskip}{1pt}
    \begin{equation}
        m_i^\ast = \arg\min_{m_i\in\mathbb{M}} F(m_i)
        \label{eq:local_goal}
    \end{equation}
\end{center}

When considering the CFL setup, the focus of the model optimization is transferred to minimizing the loss of the global model. Let $m_g = g(m_1, ..., m_N)$ be the aggregated global model after the aggregation function $g(m)$ in the central server, and $F(m_g) = \sum_{(x_{m,n},y_{m,n}) \in \mathbb{D}}f(m_g(x_{m,n});y_{m,n})$ donate the population loss function of the whole system. Then, the objective of training the global model is to find the optimal $m_g^\ast$, as defined in Equation~\eqref{eq:cfl_goal}:
\begin{center}
\setlength{\abovedisplayskip}{1pt}
    \begin{equation}
        m_g^\ast = \arg\min_{m_g\in\mathbb{M}} F(g(m_1, ..., m_N))
        \label{eq:cfl_goal}
    \end{equation}
\end{center}

However, in the DFL setup, each participant $c_i$ is only connected to a specific subset, called neighbors, $\mathbb{C}'_i$ of the entire participants $\mathbb{C}$. As a result, each participant aggregates its own local model $m_i$ with the models shared by its neighbors  $\mathbb{M}'_i$ as $m_{gi}' = g'(m_i, m_j \in \mathbb{M}'_i)$, and the corresponding population loss function for this participant changes to $F'(m_{gi}') = \sum_{(x_{m,n},y_{m,n}) \in \mathbb{D'}}f(m_{gi}'(x_{m,n});y_{m,n})$. The objective for the aggregated model in each participant, therefore, becomes to minimize the population loss function $F'(m_{gi}')$ by finding the optimal $(m_{gi}'^\ast)$ as defined in Equation~\eqref{eq:dfl_goal}:
\begin{center}
\setlength{\abovedisplayskip}{1pt}
    \begin{equation}
        m_{gi}'^\ast = \arg\min_{m_{gi}\in\mathbb{M}} F(g'(m_i, m_j \in \mathbb{M}'_i))
        \label{eq:dfl_goal}
    \end{equation}
\end{center}

It is worth noting that the count of $N'$ can vary significantly depending on the topology, ranging from 1 (representing an edge node) to $N$-1 (representing a node that is fully connected with all other nodes).

The previous problem setup assumes that the entire FL environment is secure, meaning there are no potential malicious attacks. However, in the context of a Byzantine environment, there is a fraction $\alpha$ of $N$ participants who are malicious, while the remaining $(1-\alpha)*N$ participants are benign. It can represent the set of malicious participants as $\mathcal{B}$, where $|\mathcal{B}|$ = $\alpha*N$. To enhance the resilience of the FL system, a defense mechanism $l$ is employed to identify and filter out all of the malicious participants $\mathcal{B}$, represented as $l(m_1, ..., m_N) = \mathbb{C} - \mathcal{B}$. The objective of the CFL system is then to identify the global model $m_g^\ast$ that minimizes the loss, as represented in Equation~\eqref{eq:cfl_byz_goal}:
\begin{center}
\setlength{\abovedisplayskip}{1pt}
    \begin{equation}
       m_g^\ast= \arg\min_{m_g\in\mathbb{M}} F(g(\mathbb{C} - \mathcal{B}))
        \label{eq:cfl_byz_goal}
    \end{equation}
\end{center}

In the context of DFL, when considering $\mathbb{C}'_i$ is a subset of the $\mathbb{C}$, it follows that $\mathcal{B}'_i$ is a subset of the entire set of malicious participants $\mathcal{B}$. Thus, the defense mechanism $l'$ is transformed into $l'(m_j\in \mathbb{M}'_i) = \mathbb{C}'_i - \mathcal{B}'_i$. The objective of the DFL system is then to find $(m_{gi}'^\ast)$ that minimizes$ F(g(l'(m)))$, as defined in Equation~\eqref{eq:dfl_byz_goal}:
\begin{center}
\setlength{\abovedisplayskip}{1pt}
    \begin{equation}
       m_{gi}'^\ast= \arg\min_{m_{gi}\in\mathbb{M}} F(g(\mathbb{C}'_i - \mathcal{B}'_i))
        \label{eq:dfl_byz_goal}
    \end{equation}
\end{center}
However, it should be noted that the size of $N'$ can vary from 1 to $N$-1, potentially causing $l'$ to filter out all connected participants and leading to a fragmented network. Therefore, when designing the defense mechanism $l'$ in DFL, it is crucial to prioritize both the robustness and the integrity of the DFL network to prevent network fragmentation and isolation.

Based on the previous analysis, it is evident that DFL and CFL exhibit distinct vulnerabilities due to varying network topologies. Thus, assessing the security risks associated with these diverse topologies is crucial. In the context of a DFL comprising $N$ participants, it is assumed that each participant has an average number of connected neighbors of $\overline{e}$, with $\alpha*N$ participants being malicious. Thus, for any benign participant, the number of connections with malicious participants follows a probabilistic hypergeometric distribution, as demonstrated in Equation \eqref{eq:miliciousprobics}:

\begin{equation} \label{eq:miliciousprobics}
    p(k) =   \left( \begin{array}{c} \alpha*N \\ k \end{array} \right)   \left( \begin{array}{c} N-1-\alpha*N \\ 2*\overline{e}-k \end{array}\right) / \left( \begin{array}{c} N-1 \\ 2*\overline{e} \end{array}\right)   
\end{equation}

At this point, for each participant, the mean number of malicious participants it may connect to is shown in \eqref{eq:exception}:
\begin{equation}
\overline{|\mathcal{B}'_i|} = \frac{2*\overline{e}* \alpha*N}{N -1}
\label{eq:exception}
\end{equation}

In a general sense, the risk associated with participant $c_i$ can be represented by Equation \eqref{eq:risk}:
\begin{equation}
r_i = \frac{2*\overline{e}* \alpha*N}{(N -1) * |\mathbb{C}'_i|}
\label{eq:risk}
\end{equation}
Equation \eqref{eq:risk} reveals that the security risk associated with $c_i$ is influenced by three factors: the average number of connected neighbors $\overline{e}$ of each participant in the network, the proportion $\alpha$ of potentially malicious participants in the network, and the number of direct connections $|\mathbb{C}'_i|$ that the participant has. A higher value of $\overline{e}$ and $\alpha$ increases the likelihood of participants being targeted for attacks, while a higher value of  $|\mathbb{C}'_i|$ decreases this risk. Consequently, based on the theoretical analysis, it can be insighted that: \1 the security risk of DFL is directly related to the network topology, and \2 when $\alpha$ and $N$ are fixed, increasing the number of trusted neighbors, \ie increasing $|\mathbb{C}'_i|$, could reduce the risk.
\section{\solution{} Architecture}
\label{sec:solution}

\begin{figure}[t]
    \centering
    \includegraphics[width=1\linewidth]{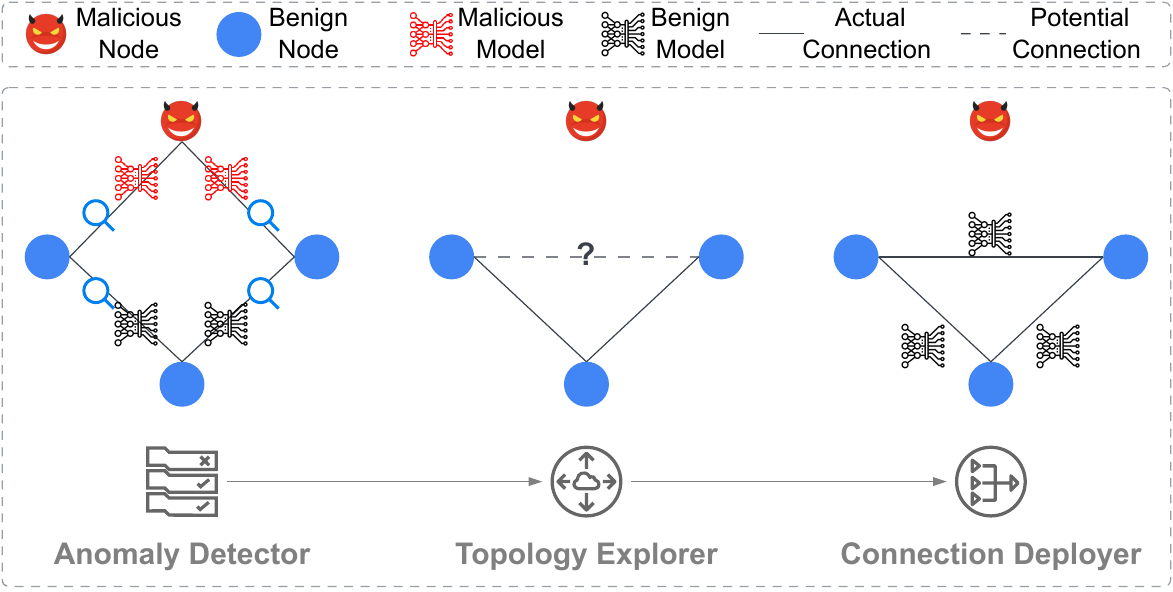}
    \caption{Overview of \solution{}}
    \label{fig:arch}
\end{figure}

Inspired by the analysis in Section \ref{sec:problemstatement}, this work proposes a three-stage aggregation protocol \solution{}. This section details the design of the \solution{} architecture that enables the DFL framework to mitigate poisoning attacks effectively. The main components of \solution{} are illustrated in \figurename~\ref{fig:arch}, comprising three modules: the anomaly detector, the topology explorer, and the connection deployer. In brief, each participant obtains the models shared by its neighbors and employs an anomaly detector to identify any abnormality within these models. If the detector distinguishes a malicious model, it activates the topology explorer component to locate other potentially connected participants. Finally, the connection deployer is responsible for establishing connections between the current participant and the recognized participants.

\subsection{Anomaly Detector}
The presence of the anomaly detector component is crucial for the effectiveness of \solution{} as it acts as a trigger for the MDT-based mechanism. According to Algorithm ~\ref{alg:ad}, each participant $c_i$ in the DFL network receives the shared models $\mathbb{M}_i'$ from its neighboring participants $\mathbb{C}_i'$. 
\begin{algorithm}[t!]
    \caption{Anomaly Detector Algorithm}\label{alg:ad}
    \begin{algorithmic}[1]
        \Require $\mathbb{M}_i'$: Neighbor models, $m_i$: Local model, $\kappa_s$: Similarity threshold.
        \State \textbf{Initialize } Triggering message $t \gets 0$
        \For{$m_j$ in $\mathbb{M}_i'$}
            \State $s_{ij} \gets CosSim(m_i, m_j)$ 
            \If{$s_{ij} >= \kappa_s$}
    	   \State  $t \gets 1$
            \EndIf
        \EndFor
        \State \Return $t$
    \end{algorithmic}
\end{algorithm}

In the initial stage, \solution{} computes the similarity between the local model $m_i$ and the received models $m_j$ in $\mathbb{M}'$. If any of these similarity scores exceeds the threshold $\kappa_s$, a triggering message $t$ is sent to the topology explorer component and the connection deployer in order to establish new connections.

\begin{algorithm}[t]
\caption{Layer-wise Cosine Similarity}\label{alg:cossim}
\begin{algorithmic}[1]
\Require $m_i$: Local model, $m_j$: Neighbor model.
\For{$l_i, l_j$ in $m_i$, $m_j$}
    \State $s_{ij} \gets s_{ij} + \frac{l_i \cdot l_j}{ \|l_i\| \|l_j\|}$ 
\EndFor
\State \Return $s_{ij}$
\end{algorithmic}
\end{algorithm}
It should be noted that the current implementation of \solution{} utilizes a layer-wise cosine similarity algorithm for detecting anomalies (as outlined in Algorithm~\ref{alg:cossim}), but alternative algorithms could also be employed for computing the similarity score.

\subsection{Topology Explorer}
The topology explorer is activated upon receiving a triggering message $t$ from the anomaly detector. As shown in Algorithm~\ref{alg:ne_r}, the node $c_i$ establishes a neighbor list $L_i$ that includes its existing neighbors $\mathbb{C}'_i$. Upon initiation of the topology explorer, the node recursively incorporates all the neighbors of the neighbor participants into its own neighbor list $L_i$ until it reaches a predefined value $\kappa_n$.

However, this approach does not take into account the potential risk of adding participants to the neighbor list $L_i$, which may lead to the inclusion of malicious models. To address this issue, the reputation of candidate participants is taken into consideration during their addition to $L_i$, only these reputation scores are higher than the reputation threshold $\kappa_r$ will be considered being added to the neighbor list $L_i$.

\begin{algorithm}[ht]
    \caption{\solution{} Neighbor Exploration with Reputation}\label{alg:ne_r}
    \begin{algorithmic}[1]
        \Require $\mathbb{C}'_i$: Neighbor participants, $\kappa_n$: Neighbor amount threshold, $m_i$: Local model, $\kappa_r$: : Reputation amount threshold.
        \State \textbf{Initialize } Neighbor list $L_i \gets \mathbb{C}'_i$
        \For{$c_j$ in $L_i$}
            \For{$c_k$ in $\mathbb{C}'_j$}
                \If{$c_k$ \textbf{not} in $L_i$ \textbf{and} $R_k >= \kappa_r$ \textbf{and} $|L_i| <\kappa_n$ }
                    \State $L_i \gets L_i + c_k$
                \EndIf
            \EndFor
        \EndFor
        \State \Return $L_i$
        \State Update the reputation score
        \For{$c_j$ in $L_i$} 
            \State $R_j \gets CosSim(m_i, m_j)$ 
            \EndFor
        \State \Return $R$
    \end{algorithmic}
\end{algorithm}

\subsection{Connection Deployer}
After acquiring the updated list of neighbors, the next step involves forming new connections between the current participant and the newly added neighbors. Then, participants exchange local models and reputation scores with others and employ the Krum~\cite{Krum:2017} aggregation function to obtain their respective global models. 

\subsection{Fedstellar}

Fedstellar is an advanced platform that streamlines the training of DFL models across multiple devices~\cite{beltran2023fedstellar}. It encompasses various significant functionalities, such as facilitating the establishment and customization of network topologies for DFL and CFL. It employs various ML/DL models and datasets to tackle various FL challenges. Besides, it provides user-friendly tools that allow for the adjustment of settings and monitoring of crucial metrics, such as resource metrics and performance metrics.
\begin{figure}[b!]
    \centering
    \includegraphics[width=1\linewidth]{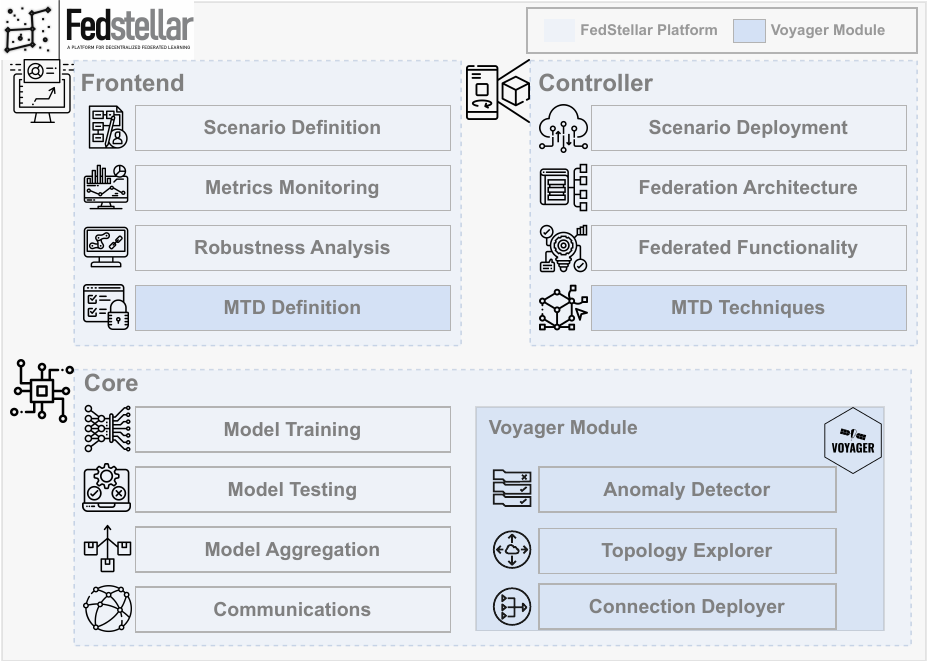}
    \caption{\solution{} Module in Fedstellar Framework}
    \label{fig:arch_in_fed}
\end{figure}

As shown in Figure 3, Fedstellar is a meticulously constructed system comprising three main elements: the frontend, the controller, and the core infrastructure.
\begin{itemize}
	\item \textbf{\textsc{frontend}}. This element facilitates the interface between the user and the system, allowing the user to customize their own FL scenario through various configurations. Additionally, the system offers real-time monitoring capabilities, such as tracking resource consumption and model performance during training process. Based on the FedStellar, \solution{} offers additional MTD selection and definition options to define MTD strategies.
	\item \textbf{\textsc{controller}}. It serves as the central center of the framework, receiving user instructions from the frontend, overseeing the entire federated scenario, choosing suitable learning algorithms and datasets, and establishing network connections to facilitate a FL procedure. The \solution{} MTD techniques are defined in this component.
	\item \textbf{\textsc{core}}. This component is incorporated into all participants of the federation and is accountable for carrying out the tasks related to FL. Its operations involve preparing data, training models, ensuring secure communication between participants, and storing the federated models. Hence, this component encompasses the main functionality of \solution{}, containing the aforementioned elements, the anomaly detector, the topology explorer, and the connection deployer.
\end{itemize}

\section{Experiments}
\label{sec:experiments}
A comprehensive assessment of the proposed \solution{} is conducted through a pool of experiments, comparing its performance to other aggregation algorithms on various datasets.
\subsection{Experiments Setup}
\subsubsection{Datasets} Three datasets and a corresponding deep learning model are chosen to evaluate the aggregation algorithms implemented for the Fedstellar framework:
\begin{itemize}
    \item \textbf{MNIST}~\cite{lecun_MNISTHandwritten_2010} consists of handwritten digits represented by 28×28 grayscale images. It compromises $60\,000$ training samples and $10\,000$ test samples. A three layers multilayer perceptron (MLP) with a linear input layer of size $784 \times 256$, a linear hidden layer of size $256 \times 128$, and a linear output layer of size $128 \times 10$ is used for classification.
    
    \item \textbf{FashionMNIST}~\cite{xiao_FashionMNISTNovel_2017} consists of $60\,000$ training samples and $10\,000$ test samples, which are 28×28 grayscale images with 10 classes.For the FashionMNIST dataset, the same MLP as for MNIST is used.
    
    \item \textbf{Cifar10}~\cite{krizhevsky_LearningMultiple_2009} consists of $50\,000$ training and $10\,000$ test images with 10 classes of 32×32 color images. A small convolutional neural network (CNN) designed for mobile applications is used for this task \cite{howard_MobileNetsEfficient_2017}. 
\end{itemize}
In a DFL network with $N$ participants, the training and testing data will be divided into $N$ equal parts for each dataset, i.e., the data distribution among the participants is considered Independent and Identically Distributed (IID). For each participant, 80\% of its training data is used for local model training and 20\% is used for model validation, \ie the validation set. In this work, $N$ is considered as 10. 



\subsubsection{Learning Configuration}
The learning process is set to a duration of 10 rounds, with each round consisting of 3 epochs. This process is conducted on the Fedstellar platform, utilizing synchronous mode. The bandwidth of the federation network is set to 1 Mbps, the loss to 0\% and the delay to 0 ms. In the Fedstellar platform, the defense mechanisms FedAvg, FLTrust, Krum, TrimmedMean, and \solution{} are implemented, evaluated, and compared. Additionally, four different network topologies (ring, star, random, and fully connected) are used in the experiments.

\subsection{Thresholds Identification}
As delineated in Chapter~\ref{sec:solution}, the algorithm employs three thresholds, specifically $\kappa_s$ for triggering the MTD, $\kappa_n$ for confirming the adequacy of the established connection, and $\kappa_r$ for assessing the suitability of the reputation score of the candidate. Therefore, this experiment employed various tests to ascertain the thresholds applicable in each scenario.

\begin{table}[t]
\centering
\caption{Cosine Similarity between Benign Models and Poisoned Models}
\label{tab:cossim}

\resizebox{1\columnwidth}{!}{%

\begin{tabular}{lll|ll|ll}

\toprule
                                    & \multicolumn{2}{c}{\textbf{MNIST}} & \multicolumn{2}{c}{\textbf{FashionMNIST}} & \multicolumn{2}{c}{\textbf{Cifar10}} \\
                                    \hline
                                                                                & Avg.    & Std.   & Avg.    & Std.    & Avg.     & Std.   \\
                                    \hline
Benign \& Benign                                                               & 0.698    & 0.018  & 0.704   & 0.008   &   0.813 &  0.003 \\
\begin{tabular}[c]{@{}l@{}}Benign \& Untargeted \\ Label Flipping\end{tabular} & 0.438    & 0.008  & 0.306   & 0.005   &  0.690  &  0.001  \\
Benign \& Model Poisoning                                                      & 0.049   & 0.001   &0.059   &0.002    & 0.440    &   0.005 \\
\bottomrule
\end{tabular}%
}
\end{table}

\tablename~\ref{tab:cossim} illustrates the cosine similarity values between the benign model and both the benign and malicious models across three different datasets. The results indicate that the benign model exhibits a higher degree of similarity with the benign model, scoring approximately 0.7 for the MNIST and FashionMNIST datasets and 0.8 for the Cifar10 dataset. Moreover, it indicates a reduced level of similarity with the model poisoned models, achieving scores below 0.5 for all three datasets. However, when considering the untargeted label flipping, the similarity score is less than 0.5 for both MNIST and FashionMNIST datasets, while approximately 0.7 for the Cifar10 dataset. As a result, this experiment sets a threshold value of 0.5 for the MTD reactor trigger $\kappa_s$ in experiments involving MNIST and FashionMNIST datasets, and $\kappa_s$=0.7 as the reactor trigger for the Cifar10 dataset. Furthermore, since the reputation score calculation utilizes the same algorithm as the anomaly detection, a reputation threshold $\kappa_r$ of 0.5 is applied for the MNIST and FashionMNIST datasets, and 0.7 for the Cifar10 dataset.

The second concern pertains to determining the optimal selection of a connection threshold $\kappa_n$ to ensure the robustness against the potential harm caused by malicious attacks. Given that the aggregation function used by \solution{} is Krum, which has a maximum tolerance for malicious equal to $\frac{N-2}{2N}$, in this case the \solution{} will try to establish more connections until $\kappa_n$. Thus, $\kappa_n$ should satisfy Equation \eqref{eq:kappa_n_condition}:

\begin{equation}
\label{eq:kappa_n_condition}
\frac{(2*\overline{e}* (1- \frac{N-2}{2N}) *N) / (N-1)}{\kappa_n} = \frac{\kappa_n-2}{2(\kappa_n)}
\end{equation}

 Thus the connection threshold $\kappa_n$ can be determined with Equation~\eqref{eq:kappa_n}, which is mainly determined by the average number of connected neighbors in the initial topology $\overline{e}$.

 \begin{equation}
\label{eq:kappa_n}
\kappa_n = min(\frac{2* \overline{e}* (N+2)}{N-1} + 2, N-1)
\end{equation}

\subsection{Poisoning Attacks Mitigation Performance}
This evaluation focuses on two types of poisoning attacks, namely untargeted label flipping and model poisoning. The evaluation of the defense mechanism is determined by the mean F1-score as $\frac{2*Precision*Recall}{Precision+Recall}$. The defense mechanism is considered more effective when the average F1-score of benign participants is higher.



\begin{figure}[t!]
\centering
\subfloat{{\includegraphics[width=9cm]{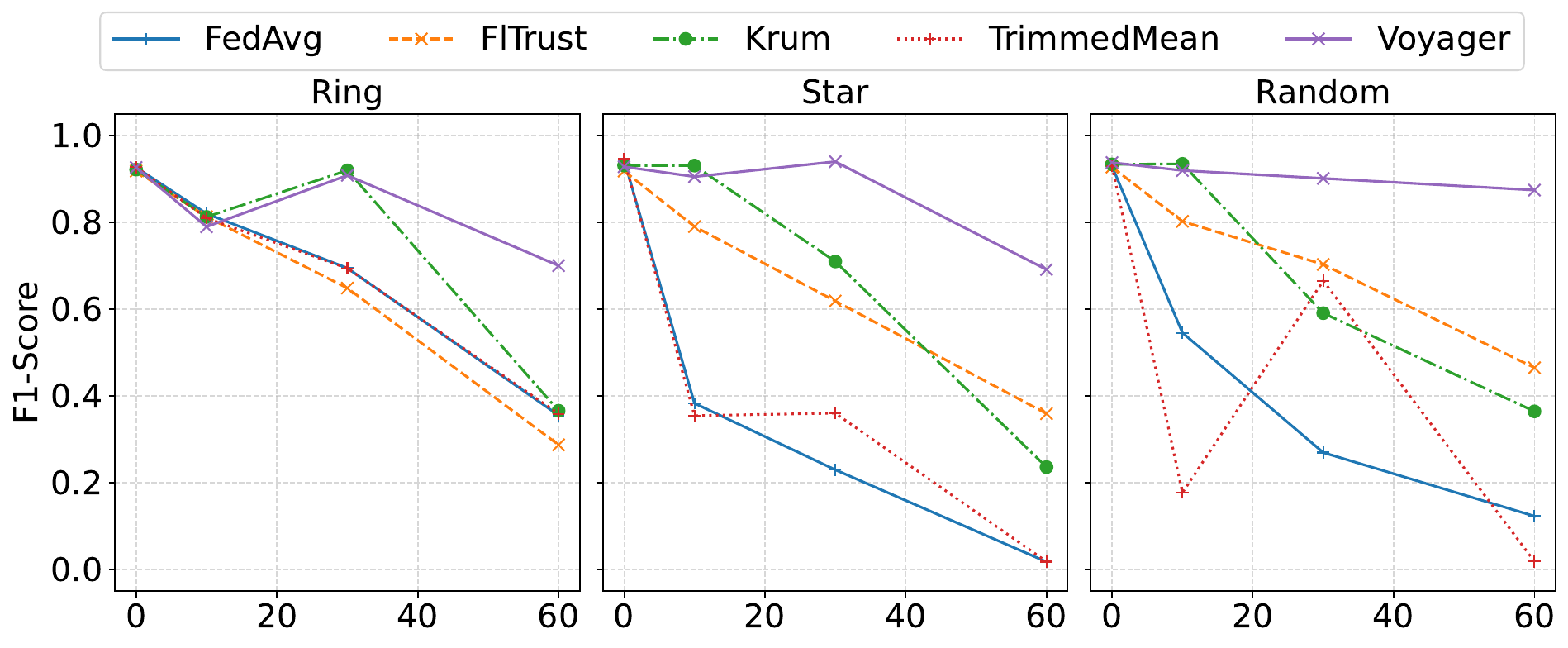}}}\\
\subfloat{{\includegraphics[width=9cm]{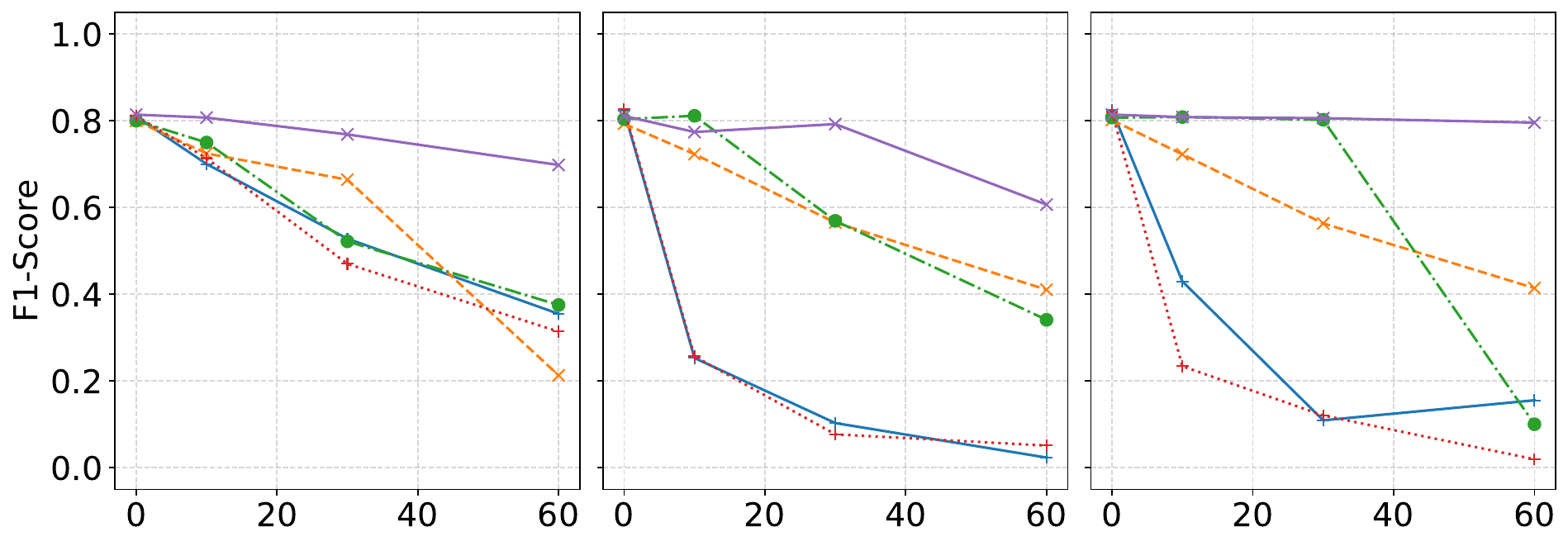}}}\\
\subfloat{{\includegraphics[width=9cm]{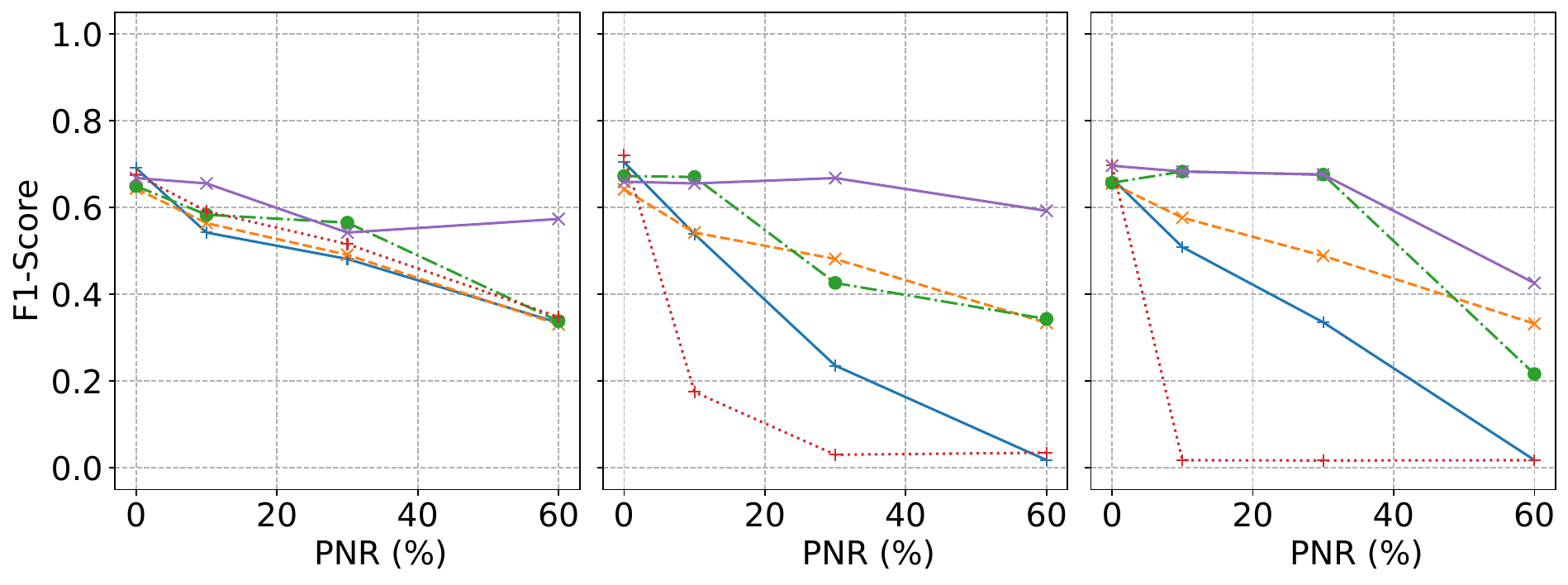}}}%
\caption{F1-score in three topologies for MNIST (top), FashionMNIST (middle), and Cifar10 (bottom) Datasets with Model Poisoning Attack}
\label{fig:modelpoisoning}
\end{figure}

\textbf{Model Poisoning.} In this attack, an adversary deliberately compromises the integrity of the model by introducing a significant amount of noise, specifically 80\% salt noise, directly into the local model after each round of the training process. The ratio of poisoned nodes (PNR) is established to 0, 10, 30, and 60. Malicious nodes are randomly selected and launched model poisoning attacks. \figurename \ref{fig:modelpoisoning} illustrates the F1-score results for the three datasets at each PNR configuration in three topologies, including ring, star, and random. The results indicate that \solution{} performed equally well as the compared algorithms in the baseline case, with a PNR of 0, across all three datasets. However, as the ratio of poisoned participants increased, Krum and TrimmedMean showed some level of resistance but became ineffective when the PNR surpassed 50\%. In contrast, \solution{} consistently outperformed the other algorithms across all three datasets and topologies, highlighting its robustness.

\begin{figure}[t!]
\centering
\subfloat{{\includegraphics[width=9cm]{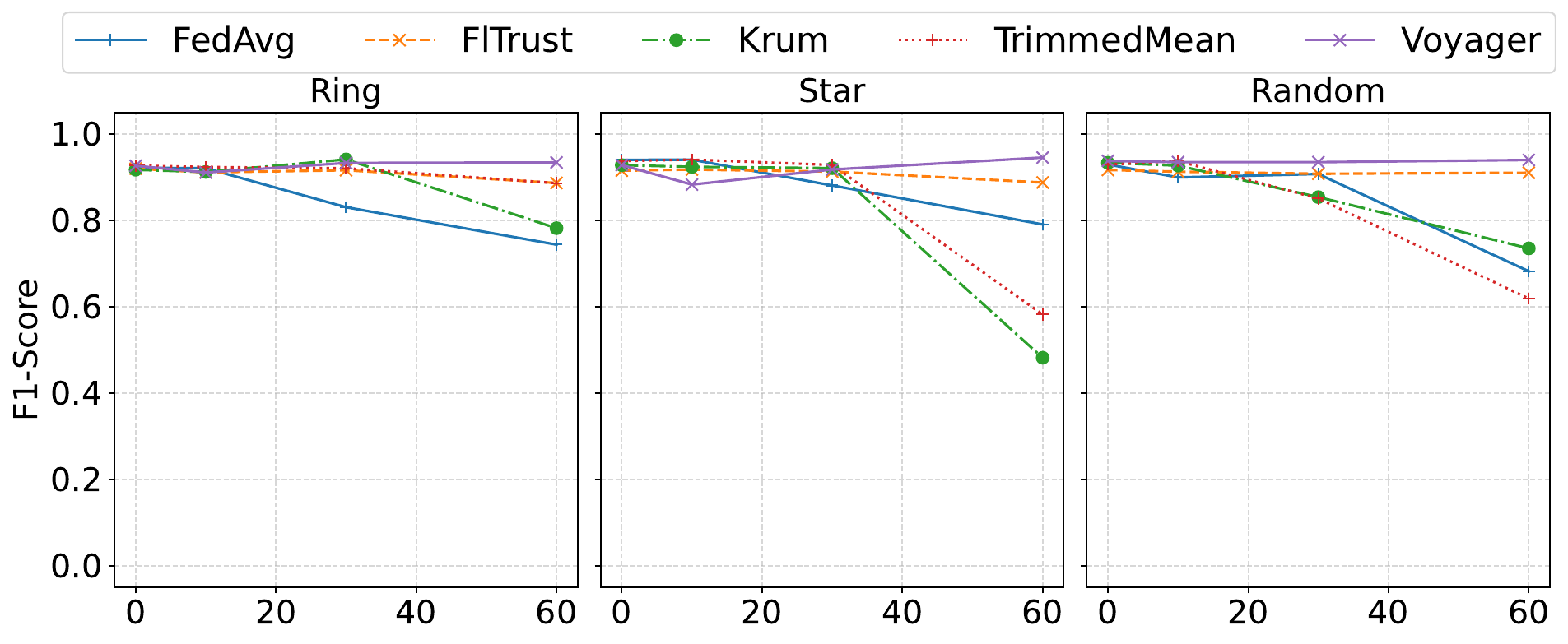}}}\\
\subfloat{{\includegraphics[width=9cm]{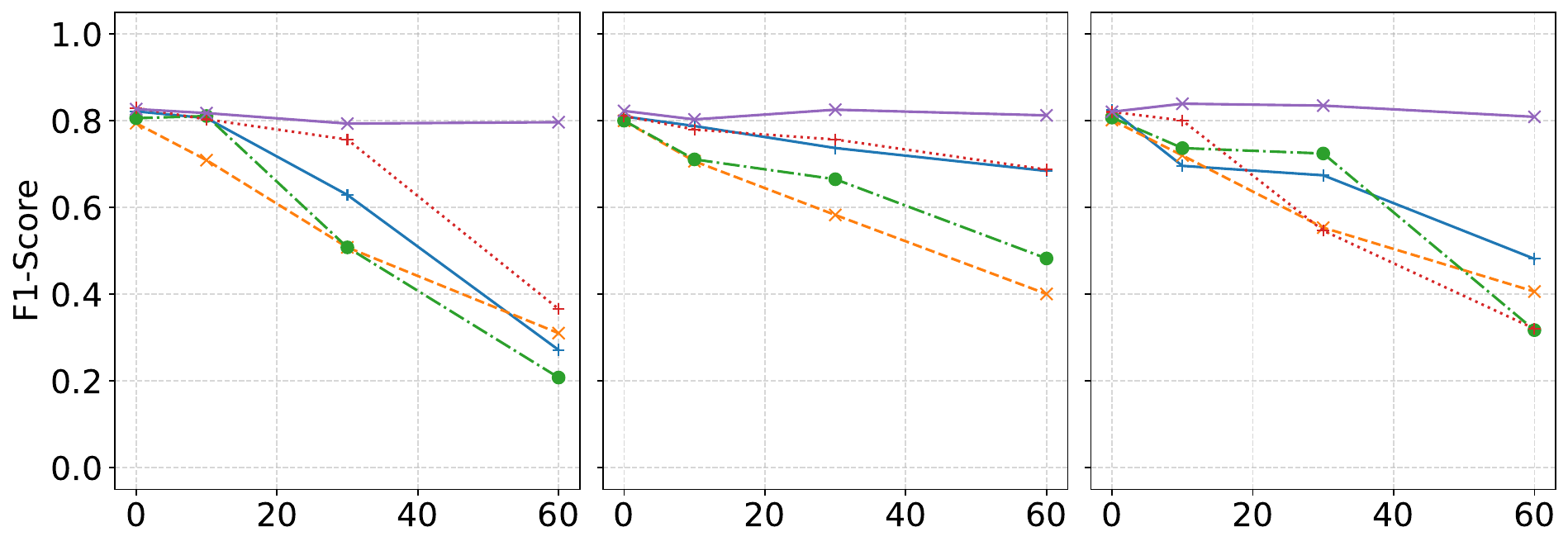}}}\\
\subfloat{{\includegraphics[width=9cm]{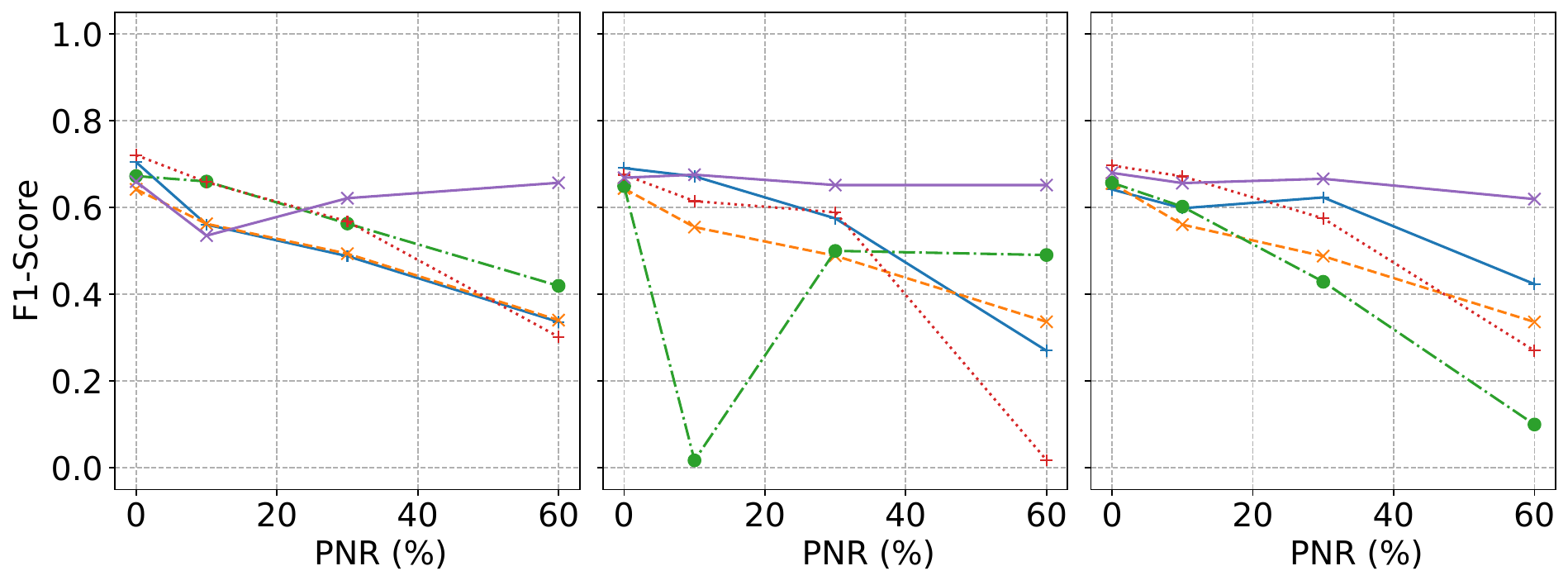}}}%
  \caption{F1-score in three topologies for MNIST (top), FashionMNIST (middle), and Cifar10 (bottom) Datasets with Untargeted Label Flipping Attack}
    \label{fig:labelflipping}
\end{figure}

\textbf{Untargeted Label Flipping.} In this attack, adversaries manipulate the local dataset by indiscriminately changing all the labels of their data samples in a random manner. The PNR exhibited variability across the datasets, with ratios of 0\%, 10\%, 30\%, and 60\%. \figurename \ref{fig:labelflipping} showcases the F1-score outcomes for all three datasets within each PNR configuration across three different topologies.

The results demonstrate that label flipping attacks pose a lower level of threat compared to model poisoning attacks. These compared algorithms exhibit resilience when the attack level is low, such as PNR at 10\%. However, similar to model poisoning attacks, the efficacy of compared algorithms declines significantly as the PNR increases. Additionally, \solution{} showcases the greatest level of resilience across all three datasets with varying topologies.

\subsection{Network Overheading and Resource Consumption}
Given that one of the strategies for \solution{} is to establish additional connections among participants, evaluating the potential network overhead and resource usage becomes essential. In the context of a DFL system consisting of $N$ nodes, where $\alpha$ fraction of $N$ nodes are benign, each node is connected to an average of $\overline{e}$ neighbors, and the size of each node's local model is assumed to be $s$ bytes. During the aggregation process, each node transmits its local model to its neighbors, resulting in network overhead as $overhead = N*\overline{e}*s$. In the case of a fully connected network, $\overline{e}$ is equal to $N-1$, thus $overhead=N*(N-1)*s$. In a ring topology, $\overline{e}$ is equal to 2, leading to $overhead=2N*s$.

With the \solution{} approach, the benign nodes break the connection with the malicious nodes while exploring for more connections, with an upper bound of $\kappa_n$ and a lower bound of 1, as shown in \eqref{eq:overhead}:

\begin{equation}
\label{eq:overhead}
\alpha*N*s<=overhead<=\alpha*N*\kappa_n*s
\end{equation}

Consider a DFL network consisting of 10 nodes with an initial toroidal topology, assuming that 80\% of the nodes are benign. If only the Krum algorithm is used, the overhead presents as  $overhead=20*s$. However, if the \solution{} algorithm is employed, it can be determined using Equation~\eqref{eq:kappa_n} that  $\kappa_n$ is equal to 7, resulting in an upper limit overhead of $overhead=56*s$. In the case of full connectivity, the overhead increases to $overhead=90*s$. 

\begin{figure}[t!]
    \centering
    \includegraphics[width=1\linewidth]{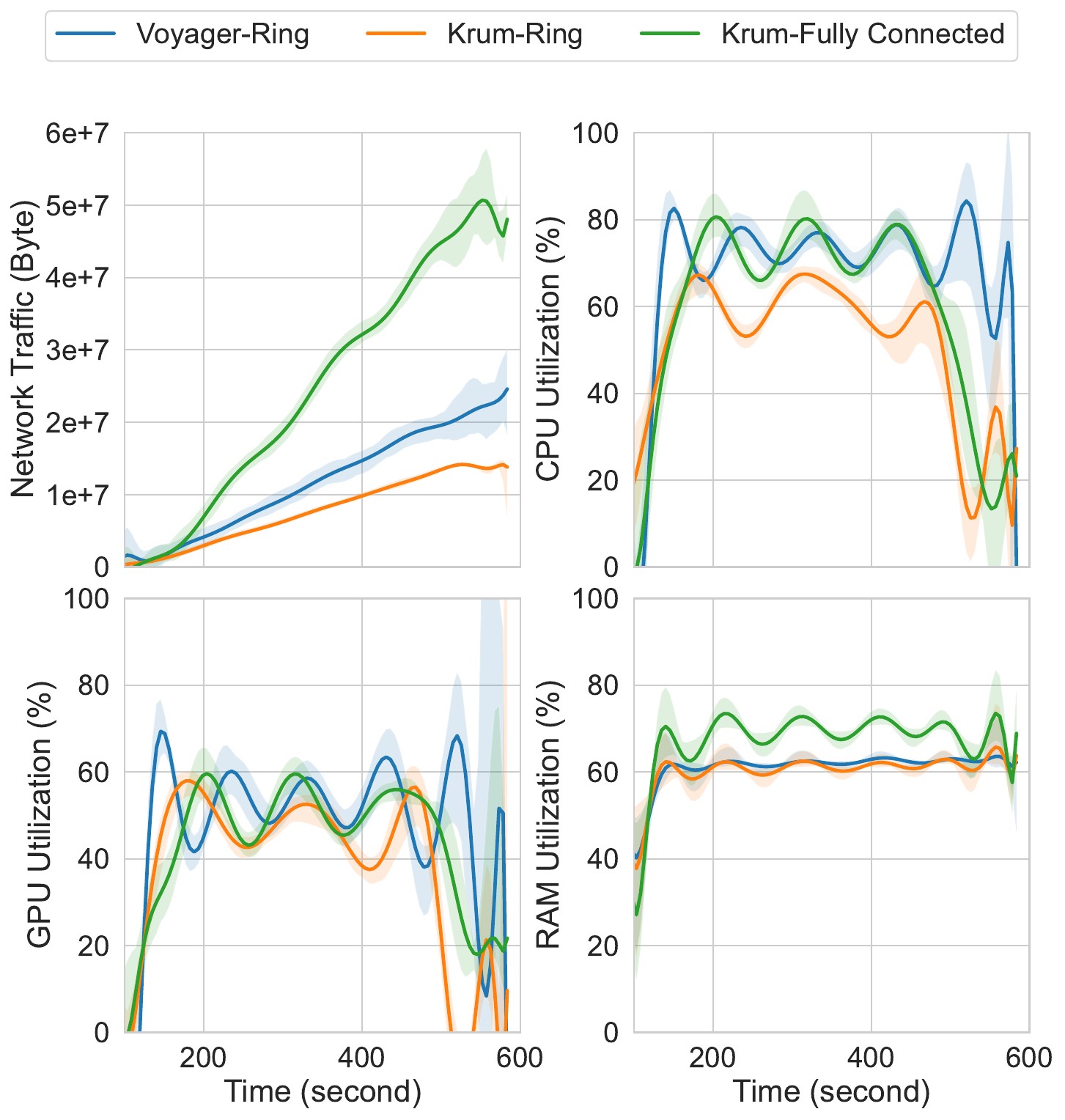}
    \caption{Network Traffic (top-left), CPU Utilization (top-right), GPU Utilization (bottom-left), and RAM Utilization (bottom-right)for Krum in Ring and Fully Connected, and Voyager in Ring with Cifar10 Dataset}
    \label{fig:network}
\end{figure}

Experiments have been carried out to compare the network traffic, CPU utilization, GPU utilization, and RAM utilization between Krum in ring and fully connected topologies, as well as \solution{} in a ring topology. These experiments use a 10-node DFL, with 80\% of the benign nodes, and the training dataset with Cifar10. The results are displayed in \figurename~\ref{fig:network}. 

In line with the previous analysis, it can be observed from top-left of \figurename~\ref{fig:network} that Krum in a ring configuration transmits the smallest quantity of data, approximately $1.5*10^7$ Byte. In comparison, \solution{} demonstrates a roughly 50\% increase in the amount of data transmitted, totaling around $2.3*10^7$ Byte, which is much less than the $5.1*10^7$ bytes transmitted by Krum in a fully connected setup. Regarding CPU utilization and GPU utilization, there is no notable distinction among the three approaches, and in general, \solution{} does not impose a substantial computational load on the system. However, in terms of RAM utilization, the Kurm in fully connected setup necessitates larger data transfers and additional buffers, resulting in a significantly higher initial RAM usage compared to \solution{} and Krum in ring. The results show that although \solution{} introduces additional network overhead, it remains within an acceptable range.


\section{Summary and Future Work}
\label{sec:conclusions}
This work proposed \solution{}, a novel defense strategy that utilizes a reactive MTD-based mechanism to mitigate poisoning attacks. \solution{} leverages the knowledge of the network structure and availability of local data in DFL to showcase the adaptability of effective defense strategies. Through extensive evaluations on different datasets such as MNIST, FMNIST, and CIFAR10, using various attack configurations, \solution{} has demonstrated its effectiveness across different network topologies. In comparison to other robust aggregation algorithms, the proposed \solution{} in this work has consistently proven to be a reliable defense against poisoning attacks, without noticeably increasing the network traffic and resource usage.

In the future, proactive strategies for MTD will be considered and implemented. Furthermore, this study only focused on analyzing datasets in an IID setting. Therefore, there are intentions to assess the performance of \solution{} in non-IID scenarios through diverse experiments.

\section*{Acknowledgments}
This work has been partially supported by \textit{(a)} the Swiss Federal Office for Defense Procurement (armasuisse) with the CyberForce project (CYD-C-2020003) and \textit{(b)} the University of Zürich UZH.

\bibliographystyle{IEEEtran}  
\balance
\bibliography{references}

\end{document}